\documentclass[%
 reprint,
 amsmath,amssymb,
 aps,
]{revtex4-1}

\usepackage{lipsum}
\usepackage{balance}
\usepackage{hyperref}
\usepackage{appendix}
\usepackage[user,titleref]{zref}
\usepackage[nameinlink,capitalize]{cleveref}
\usepackage{braket}
\usepackage{natbib}
\usepackage{subfigure}
\usepackage{tikz}
\usepackage{graphicx}% Include figure files
\usepackage{dcolumn}% Align table columns on decimal point
\usepackage{bm}% bold math
\begin{document}

\preprint{APS/123-QED}

\title{Phenomenological inclusion of alternative dispersion relations to the Teukolsky equation and its application to bounding the graviton mass with gravitational-wave measurements}% Force line breaks with \\

\author{Adrian Ka-Wai Chung}
\email{kwchung@phy.cuhk.edu.hk}
\affiliation{%
 Department of Physics, The Chinese University of Hong Kong, Shatin, N.T., Hong Kong
}%

\author{Tjonnie Guang Feng Li}
\email{tgfli@cuhk.edu.hk}
\affiliation{%
 Department of Physics, The Chinese University of Hong Kong, Shatin, N.T., Hong Kong
}%

\begin{abstract}
	Existing constraints on the graviton mass from gravitational-wave detections rely on the phase difference developed between different frequencies during the propagation.
	Effects on the quasinormal-mode frequencies of the black-hole ringdown due to the graviton mass are often ignored.
	While perturbation theories of black holes have been well developed in the context of general relativity, this is not the case for modified gravity theories.
	We propose a phenomenological modification to the Teukolsky equation of perturbed black holes to include the dispersion relation due to a gravitational field of nonzero mass.
	Solving this modified Teukolsky equation by logarithmic perturbation theory, we compute the shift of the quasinormal-mode frequencies due to the presence of a graviton mass.
	This hypothetical shift can be used to constrain the graviton mass with ringdown signals, either standalone or in conjunction with the phase difference accumulated due to the wave propagation.
	We estimate that constraints on the graviton mass of $m_g \lesssim 10^{-15}\,\textrm{eV}$ can be put with a detection of the ringdown signal alone by second generation gravitational-wave detectors.
\end{abstract}

\pacs{Valid PACS appear here}% PACS, the Physics and Astronomy
\maketitle

%\tableofcontents 

\section{Introduction}

Direct detections of gravitational waves emitted by merging binaries with the Advanced LIGO and Virgo detectors \cite{LIGO_01, LIGO_02, LIGO_03, LIGO_04, LIGO_05, LIGO_06} have provided opportunities to test general relativity \cite{LIGO_07, LIGO_08}.
The dispersion relation of gravitational waves and the graviton mass are common aspects of these tests.
According to general relativity, gravitational waves are local Lorentz invariants.
Therefore, gravitons should have zero mass and obey the dispersion relation of $ \omega = k $.
Existing gravitational-wave detections show no deviations from this corollary of general relativity \cite{LIGO_07, LIGO_08}.

Existing constraints on the graviton mass from gravitational-wave detections rely on the weak-field propagation of gravitational waves \cite{Wills_01, Wills_02, Ajith_01, Samajdar_01}.
A massive graviton is expected to alter the dispersion relation of gravitational-wave to $ \omega^2 - k^2 = m_g^2 $, where $ m_g $ is the graviton mass.
Gravitational waves of different frequencies following this dispersion relation travel at different propagation velocities.
Consequently, a phase difference develops between different frequencies as gravitational waves propagate.
The absence of this phase difference allows one to put constraints on $ m_g $ up to the reciprocal of the Compton wavelength of the propagation distance.
However, this constraint is limited to the weak-field propagation of gravitational waves.
The behavior of the alternative dispersion relations in the strong field remains untested.

While the effects of modified gravity theories \cite{LVTheory_01, LVTheory_02, LVTheory_03, LVTheory_04, LVTheory_05} on the inspiral of merging binaries have been well studied (see e.g. Ref.~\cite{Yunes_01} and references therein), studies of these effects on the post-merger and ringdown are mostly numerical simulations \cite{Chern_Simons_01, Helvi_01, Helvi_02, Helvi_03, RD_Chern-Simons, RD_f_R, RD_Gauss_Bonnet, RD_Einstein_Maxwell_Dilaton, RD_Einstein-dilaton-Gauss-Bonnet}. 
Due to the computational complexity, it is impossible to directly apply numerical simulations to parameter estimation of gravitational-wave signals.
Even though black holes in the ringdown stage can be described by perturbation theories, these have not been fully developed for alternative theories.
There has been extensive studies of massive perturbation fields to black holes \cite{Massive_BH_Perturbation_01, Massive_BH_Perturbation_02, Massive_BH_Perturbation_03, Massive_Gravity_01, Massive_Scalar_Perturbation_01, Massive_Scalar_Perturbation_02, Massive_Vector_Perturbation_01, Perimeter_01, Berti_04, Berti_05}.
However, these studies are confined to either Schwarzschild or slowly spinning black holes, because the perturbation equations are generally not separable in the Kerr metric.

The goal of this paper is to introduce a phenomenological modification to the Teukolsky equation of perturbed black holes to include an alternative dispersion relation of gravitational waves.
The proposed modification keeps the perturbation equation separable, which opens up the possibility to consider black holes of arbitrary spins.
In particular, we calculate the shift in quasinormal modes due to the graviton mass.
This allows us to probe the graviton mass using gravitational waves from the black-hole ringdown.
This paper is organized as follow:
Section II outlines a proposed phenomenological modification to the Teukolsky equation to account for a modified dispersion relation.
Section III discusses the parameter estimation of the graviton mass from the black-hole ringdown.
In section VI, we discuss the implications of our study.

Throughout this paper, we will work in units of $ c = \hbar = 1 $ for $ m_g $.
Therefore, $ m_g $ shares the same dimensionality with frequency $[\textrm{s}^{-1}]$. 
$ m_g = 1\textrm{s}^{-1} \approx 4 \times 10^{-15}\,\textrm{eV}$. 
The signature of $ g_{\mu \nu} = (+, -, -, -) $ is assumed.

\section{METHOD}

For a Kerr black hole of mass $ M $ and angular momentum $ M a $, scalar, vector and tensor perturbations obey the Teukolsky equation \cite{Teukolsky_01_PRL, Teukolsky_02_ApJ, Teukolsky_03_ApJ, Teukolsky_04_ApJ}:
\begin{equation}\label{eq:TE}
\mathcal{L} \psi = 4 \pi T, 
\end{equation}
where $ \mathcal{L} $ is a linear differential operator involving at most the second order derivatives with respect to the Boyer-Lindquist coordinates, $ (t, r, \theta, \phi) $ and $ T $ is the source term of the black-hole perturbation.
$ \mathcal{L} $ also depends on $ s $, the spin weight of perturbation field (See \cref{eq:MTE_massive_gravity} for the explicit form), where $ s = 0 $ for scalar fields, $ s = \pm 1 $ for vector fields and $ s = \pm 2 $ for gravitational fields.
These perturbation fields to the metric are massless.
This is manifested by the following properties of the Teukolsky equation:
(i) It reduces to $ \Sigma g^{\mu \nu} \nabla_{\mu} \nabla_{\nu} \psi = 0 $, where $ \Sigma = r^2 + a^2 \cos^2 \theta $, when scalar perturbations in vacuum are considered.
This is the equation of motion of a massless scalar field in curved spacetime.
(ii) For all types of perturbation, $ r \rightarrow \infty \Rightarrow \mathcal{L} \psi = 0 \rightarrow \Sigma (\partial_t^2 - \partial_r^2) \psi = 0 $, which is the wave equation of a massless field.
These suggest that perturbation to black holes in modified gravity theories requires a separate treatment.

To incorporate alternative dispersion relations of gravitational waves suggested by different modified gravity theories, for example, massive gravity \cite{MG_Pauli, MG_02, MG_03, MG_04, MG_05, MG_06}, we proposed a phenomenological modification, based on the following observations.
Consider a general dispersion relation of $ \omega^2 - k^2 = \mathcal{D} (\theta, \phi; k) $, which is defined relative to an observer in the weak-field regime ($r \rightarrow + \infty $).
We demand that $ \psi $ obeys a wave equation in the form of $ (\partial_t^2 - \partial_r^2 + \mathcal{D}) \psi = 0$ in this weak-field regime.
Compared the weak-field Teukolsky equation, an extra term of $ \mathcal{D} \Sigma \psi $ is needed on the lhs of \cref{eq:TE}.
Thus, a possible extension of the Teukolsky equation which includes the modified dispersion is given by
\begin{equation}\label{eq:MTE}
\mathcal{L} \psi + \mathcal{D} \Sigma \psi = 0.
\end{equation}

As a corollary, for a perturbation field of mass $ m $, spin weight $ s $ and dispersion term $ \mathcal{D} = m^2 $, \cref{eq:MTE} becomes $ \mathcal{L} \psi + m^2 \Sigma \psi = 0 $. 
Explicitly, in units of $ c = G = 1 $, we have
\begin{widetext}
\begin{equation}\label{eq:MTE_massive_gravity}
\begin{split}
&\Bigg( \frac{(r^2 + a^2)^2}{\Delta} - a^2 \sin^2 \theta \Bigg) \frac{\partial^2 \psi}{\partial t^2} + \frac{4 Mar}{\Delta} \frac{\partial^2 \psi}{\partial t \partial \phi} + \Bigg( \frac{a^2}{\Delta} - \frac{1}{\sin^2 \theta} \Bigg) \frac{\partial^2 \psi}{\partial \phi^2} - \Delta^{-s} \frac{\partial }{\partial r} \Big( \Delta^{s+1} \frac{\partial \psi}{\partial r}\Big) - \frac{1}{\sin \theta} \frac{\partial }{\partial \theta} \Big( \sin \theta \frac{\partial \psi}{\partial \theta} \Big) \\
& - 2s\Bigg( \frac{a(r-M)}{\Delta} + i \frac{\cos \theta}{\sin^2 \theta} \Bigg)\frac{\partial \psi}{\partial \phi} - 2s \Bigg(\frac{M(r^2 - a^2)}{\Delta} - r - ia \cos \theta \Bigg) \frac{\partial \psi}{\partial t} +(s^2 \cot^2 \theta - s) \psi + m^2 (r^2 + a^2 \cos^2 \theta)\psi = 0 .  
\end{split}
\end{equation}
\end{widetext}
where $ \Delta = (r - r_-)(r - r_+) $ and $ r_{\pm} = M \pm \sqrt{M^2 - a^2} $ are the outer and inner horizons of the rotating black hole.
For a massive scalar field ($s=0$), \cref{eq:MTE_massive_gravity} reduces to $ \Sigma (g^{\mu \nu} \nabla_{\mu} \nabla_{\nu} \psi + m^2 ) \psi = 0 $, which is the Klein-Gordon equation in covariant form \cite{Klein_Gordon_equation_01}.
For gravitational perturbations with a non-zero field mass, we take $ s = -2 $ and $ m = m_g $, where $ m_g $ is the graviton mass, in \cref{eq:MTE_massive_gravity}.
In general, $m^2 \Sigma \psi $ depends on both $ r $ and $ \theta $ due to frame dragging around a rotating black hole.
This term changes both the amplitude and the angular dependence of different quasinormal modes. 
We concentrate only on the effect on quasinormal mode frequencies and ignore all the other effects by the graviton mass, because the former is dominant in terms of detectability.
Other effects due to the graviton mass, for example, emergence of additional quasinormal modes, polarizations and breaking of isospectrality (see e.g. \cite{Massive_Gravity_01}), are ignored. 
For these reasons, $ m_g $ in \cref{eq:MTE_massive_gravity} is actually a phenomenological graviton mass. 

The effects of the mass of the gravitational field on quasinormal-mode frequencies can be calculated by solving \cref{eq:MTE_massive_gravity} with $ s = -2 $.
By separation of variables, $ \psi = R(r) S(\theta) e^{i m \phi + i \omega t} $, where $ S(\theta) $ is the spheroidal function that depends on $ \theta $, the angle between line of sight and the spin of the black hole.
Let $ u = \Delta^{s/2} (r^2 + a^2)^{1/2} R $.
$u$ then satisfies a Schr\"{o}ndinger-like equation \cite{Nakamura_01}, 
\begin{equation}\label{eq:SE}
\frac{\partial^2 u}{\partial r_*^2} + \Bigg( \omega^2 - V(r) - m_g^2 \frac{r^2 \Delta}{(r^2 + a^2)^2} \Bigg) u = 0, 
\end{equation}
where $ V $ is the effective potential generated by the background geometry of the black hole and $ r_* $ is the tortoise coordinate, defined by $ \frac{d}{d r_*} = \frac{\Delta}{r^2+a^2} \frac{d}{dr}$.
When $ m_g = 0 $, \cref{eq:SE} reduces to the radial part of \cref{eq:TE} for $ T = 0 $. 
Given that recent constraints on $ m_g $ indicate that it is approximately massless, we assume $ \omega^2 >> m_g^2 $ \cite{LIGO_07, LIGO_08}.
With this assumption, quasinormal-mode frequencies of black holes described by \cref{eq:SE} can be approximated by perturbation theory. 
Following the recipe of logarithmic perturbation theory (LPT) \cite{LPT_01, LPT_02, LPT_03}, we expand the complex frequencies of the overtone $ nlm $ as $ \tilde{\omega}_{nlm} =  \tilde{\omega}_{nlm}^{(0)} + \tilde{\omega}_{nlm}^{(1)} $, where $ \tilde{\omega}_{nlm}^{(0)} $ is the unperturbed frequency and $ \tilde{\omega}_{nlm}^{(1)} $ is the leading order shift due to the $m_g^2$ term, which is of order $ \mathcal{O} (m_g^2) $.
The perturbed quasinormal-mode frequencies are given by
\begin{equation}\label{eq:freq_shift}
\begin{split}
\tilde{\omega}^{(1)}_{nlm} &\approx \frac{m_g^2}{2 \tilde{\omega}^{(0)}_{nlm}}, \\
\tilde{\omega}_{nlm} &\approx \tilde{\omega}^{(0)}_{nlm} + \frac{m_g^2}{2 \tilde{\omega}^{(0)}_{nlm}}.  
\end{split}
\end{equation}
We refer readers for details of the LPT calculation to the appendix.
The (real) frequency and lifetime of the $nlm$ overtone are $ \omega_{nlm} = \omega^{\rm Re}_{nlm} = \text{Re} \tilde{\omega}_{nlm} $ and $ \tau_{nlm} = 1 / \omega^{\rm Im}_{nlm} = 1 / \text{Im} \tilde{\omega}_{nlm} $, respectively.

\begin{figure}[htp!]
  \centering
  \includegraphics[width=\columnwidth]{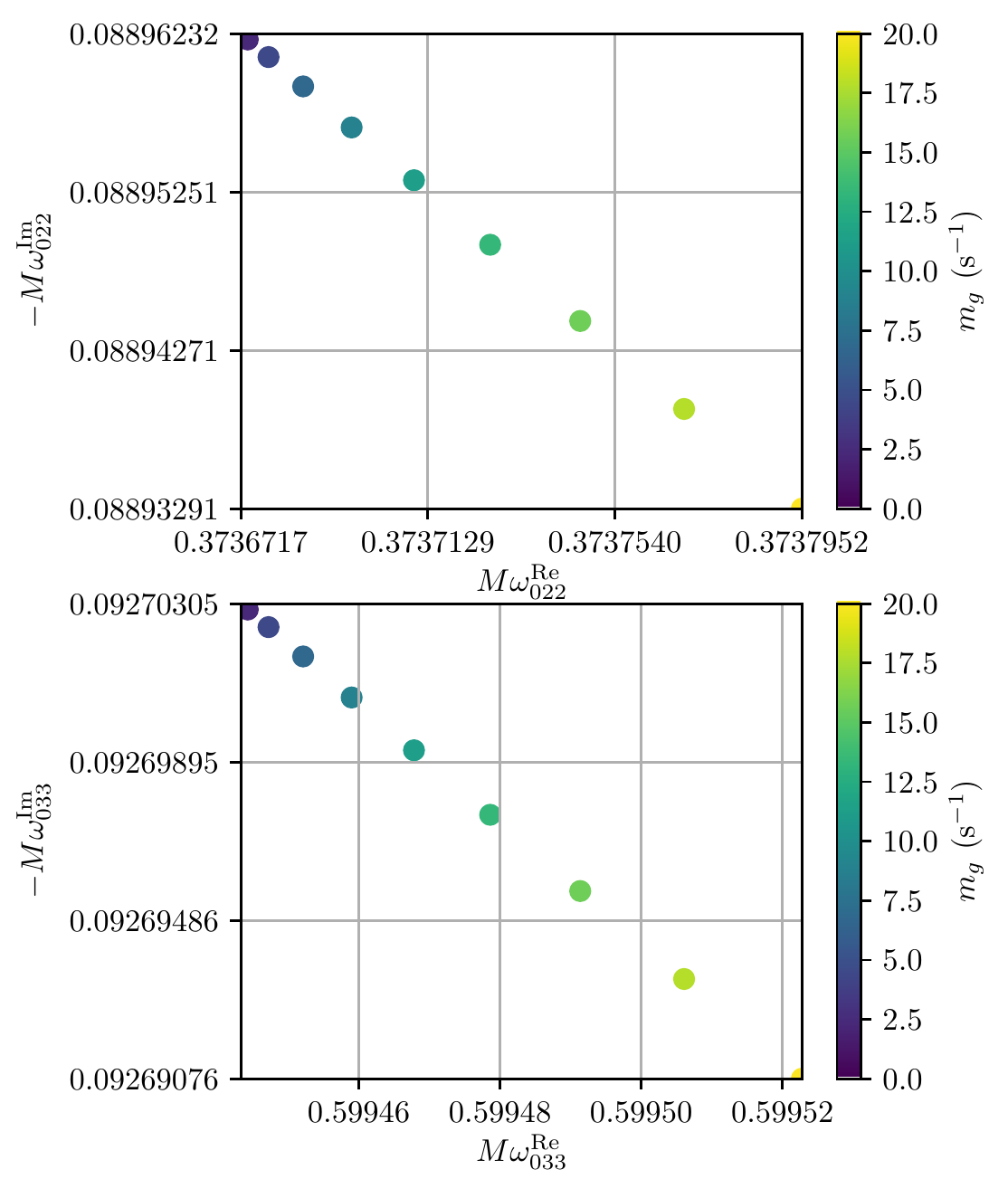}
  \caption{Quasinormal mode frequencies of 022 (top panel) and 033 (bottom panel) overtone of a spinless black hole of mass $ 100 M_{\odot}$ with $ m_g = 0, 2, 4, ...., 20 \,\textrm{s}^{-1}$. The point of the smallest $ \omega^{\rm Re}_{nlm} $ and the largest $ \omega^{\rm Im}_{nlm} $ corresponds to $ m_g = 0 $. Both frequencies and lifetimes are enhanced by mass of gravitational field.}
  \label{fig:complex_plane}
\end{figure}

\cref{fig:complex_plane} plots the complex quasinormal-mode frequencies of the $022$ (top panel) and $033$ (bottom panel) overtones of a spinless black hole of $ 100 M_{\odot} $ with different values of the gravitation mass $ m_g = 0, 2, 4, ..., 20\,\textrm{s}^{-1}$.
%Both frequencies and lifetimes of various overtones of quasinormal modes are increased due to the presence of the graviton mass.
%The $y$-axes plot $ M \omega^{\rm Im}_{nlm} $ and the $x$-axes plot $ M \omega^{\rm Re}_{nlm} $.
%The smallest of $ \omega^{\rm Re}_{nlm} $ and the largest of $ \omega^{\rm Im}_{nlm} $ of both overtones correspond to $ m_g = 0 $.
As $ m_g $ increases, complex frequencies of both modes follow a trajectory to the bottom right of complex planes, which implies both frequencies and lifetimes are enhanced by $ m_g $.
Our finding is consistent with the existing results for massive gravitational fields \cite{Massive_Gravity_01}, vector fields \cite{Massive_Vector_Perturbation_01} and scalar fields \cite{Massive_Scalar_Perturbation_01, Massive_Scalar_Perturbation_02}.
\cref{eq:freq_shift} provides an analytical expression for computing quasinormal-mode frequencies of black holes in massive gravitational fields, which can be conveniently used for parameter estimation efforts.

The mass of the gravitational field changes the quasinormal mode frequencies due to two reasons.
Firstly, when $ m_g > 0 $, gravitational waves propagate at speed slower than the speed of light $ v_g = 1 - \frac{m_g^2}{2 \omega}$ \cite{Wills_01}. 
Quasinormal-mode frequencies in turn scale as $ v_g / M $.
Therefore, a change of $ v_g $ due to a massive gravitational field alters the quasinormal-mode frequencies of black holes.
Secondly, $m_g \neq 0$ develops an extra effective potential of gravitational perturbation around black holes.
Contribution to the effective potential by the graviton mass selects different characteristic frequencies of gravitational waves which are able to propagate toward spatial infinity (leak through the potential) \cite{QNM_Rev_01}.

We implement these changes to the quasinormal-mode frequencies into a multi-mode ringdown waveform model that is calibrated against numerical simulations \cite{Nikhef_01}. 
In particular, for a Boyer-Lindquist observer at distance $ d_L $ from the black hole, the ringdown waveform of the $ nlm $ overtone looks like
\begin{equation}\label{eq:RDTDWF}
h_{nlm}(d_L, \theta, \phi, t) = \frac{M}{d_L} A_{nlm} (\eta, \chi) S_{lm}(\theta, \phi) e^{+i \tilde{\omega}_{nlm} t},
\end{equation}
where $ \theta $ is the angle between the line of sight and spin of the black hole, $ \phi$ is the azimuthal angle, $ A_{nlm} $ is the amplitude of $ nlm $ the overtone, which is a function of symmetric mass ratio $ \eta $, spins of the parental black holes $ \chi $ and $ S_{lm} $ is the spin-weighted spheroidal wave functions \cite{Nikhef_01}. 
The complex frequencies $ \tilde{\omega}_{nlm} $ of \cref{eq:RDTDWF} are given by \cref{eq:freq_shift}.

\begin{figure}[h!]
  \centering
  \includegraphics[width=\columnwidth]{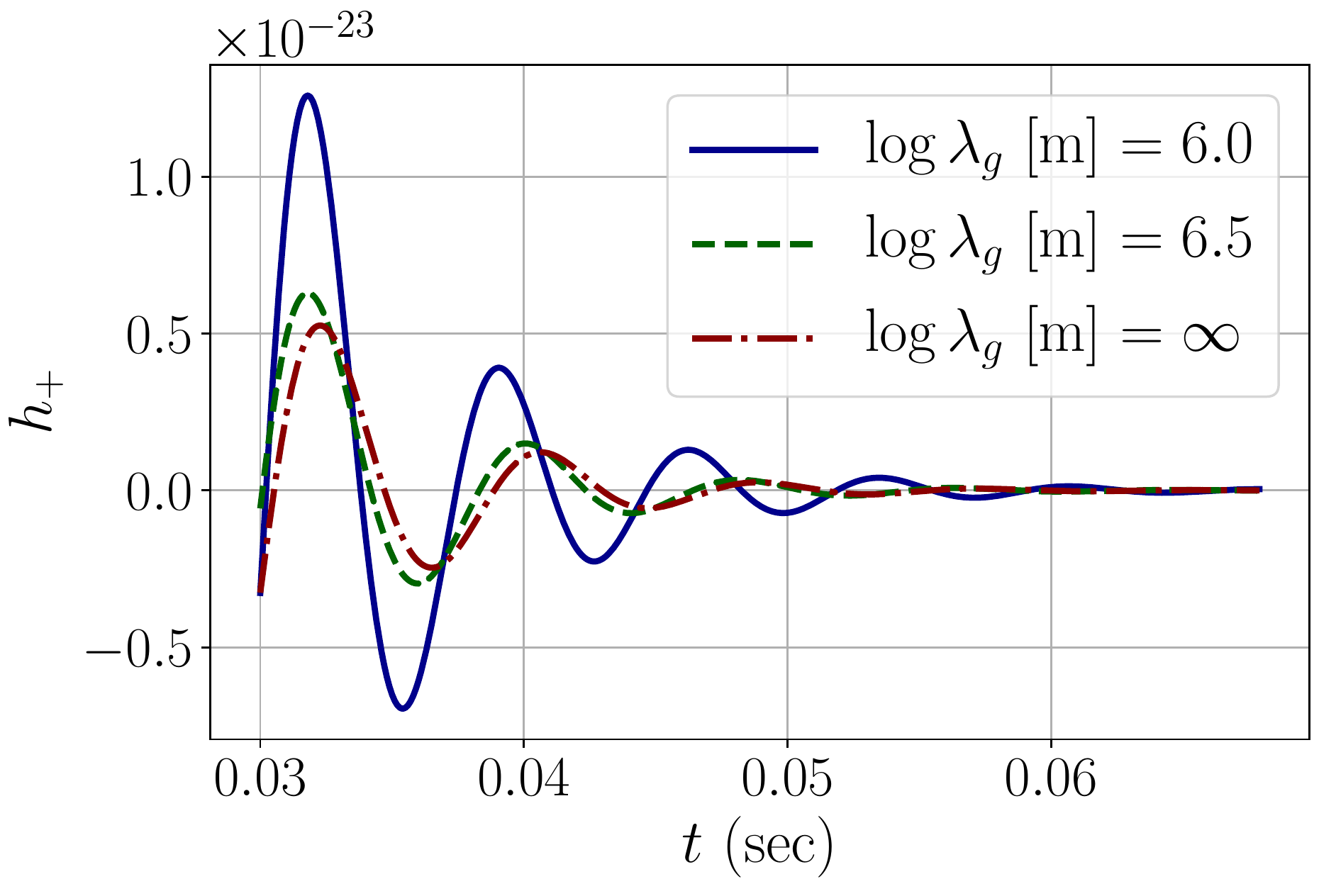}
  \caption{(color online). The plus mode time domain ringdown waveforms emitted by a black hole of 100 $ M_{\odot} $ at 400\,Mpc for $ \log  \lambda_g = $ 6.0 (blue), 6.5 (green), and $ \log \lambda_g = \infty $ ($ m_g = 0 $) (red). As $ \log \lambda $ approaches 7, waveforms of $ m_g > 0 $ and $ m_g = 0 $ overlap almost completely. Major overtones of $ nl|m| = $ 022, 122, 033, 133, 044, 055, 021, 032 and 034 are included in the plots.}
  \label{fig:RDTDWF}
\end{figure}

\cref{fig:RDTDWF} shows the time domain ringdown waveforms for several values of $ \lambda_g $, emitted by black hole of  100\,$ M_{\odot} $ at 400\,Mpc away when $\theta = \pi / 2$.
Overtones of the $ nl|m| = $ 022, 122, 033, 133, 044, 055, 021, 032 and 034 modes are included in the plots.
These are the dominant modes in the ringdown stage found in numerical simulations \cite{QNM_01, QNM_02}.
The phase difference due to the propagation between different frequencies are ignored.
Both the frequencies and lifetimes of ringdown are increased compared with that of shorter wavelengths $ \lambda_g $, which corresponds to more massive gravitons.
As $ \log \lambda_g $ approaches $ \sim 7.0$, waveforms of $ m_g > 0 $ and $ m_g = 0 $ overlap almost completely.
In conclusion, the graviton mass increase both the frequencies ($ \tilde{\omega}^{\text{Re}}_{nlm} $) and lifetimes ($ 1 / \tilde{\omega}^{\text{Im}}_{nlm} $).
%The shifts of quasinormal mode frequencies allows us to estimate the graviton mass.

\section{PARAMETER ESTIMATION}
The shifts of quasinormal mode frequencies due to the graviton mass [\cref{eq:freq_shift}] allow us to estimate or constrain the graviton mass $ m_g $.
We implement the modified waveforms [\cref{eq:RDTDWF}] to LALInference, the standard parameter estimation software used by the LIGO-Virgo Collaboration \cite{LALInference}.
An extra free parameter of $ \log \lambda_g $ is added into the waveforms with all aforementioned overtones included, where $ \lambda_g $ is the Compton wavelength of the phenomenological graviton mass, 
\begin{equation}
\lambda_g = \frac{h}{mc} = \frac{1}{m}. 
\end{equation}
We simulate a set of ringdown signals with $m_g = 0 $ by black holes of masses between $ M \in [10, 290] M_{\odot} $ at a luminosity distance of $d_L = 400\,\textrm{Mpc}$, roughly the distances of the first two detected events \cite{LIGO_01, LIGO_02}, in stationary Gaussian noise.
%Aforementioned overtones are included in the parameter estimation.
The nested sampling algorithm implementation within \textsf{LALInference} was used to infer $ \log \lambda_g $.
The prior of $ \log \lambda_g $ is set to be uniform over $[0, 30]$.

\begin{figure}[htp!]
  \centering
  \includegraphics[width=\columnwidth]{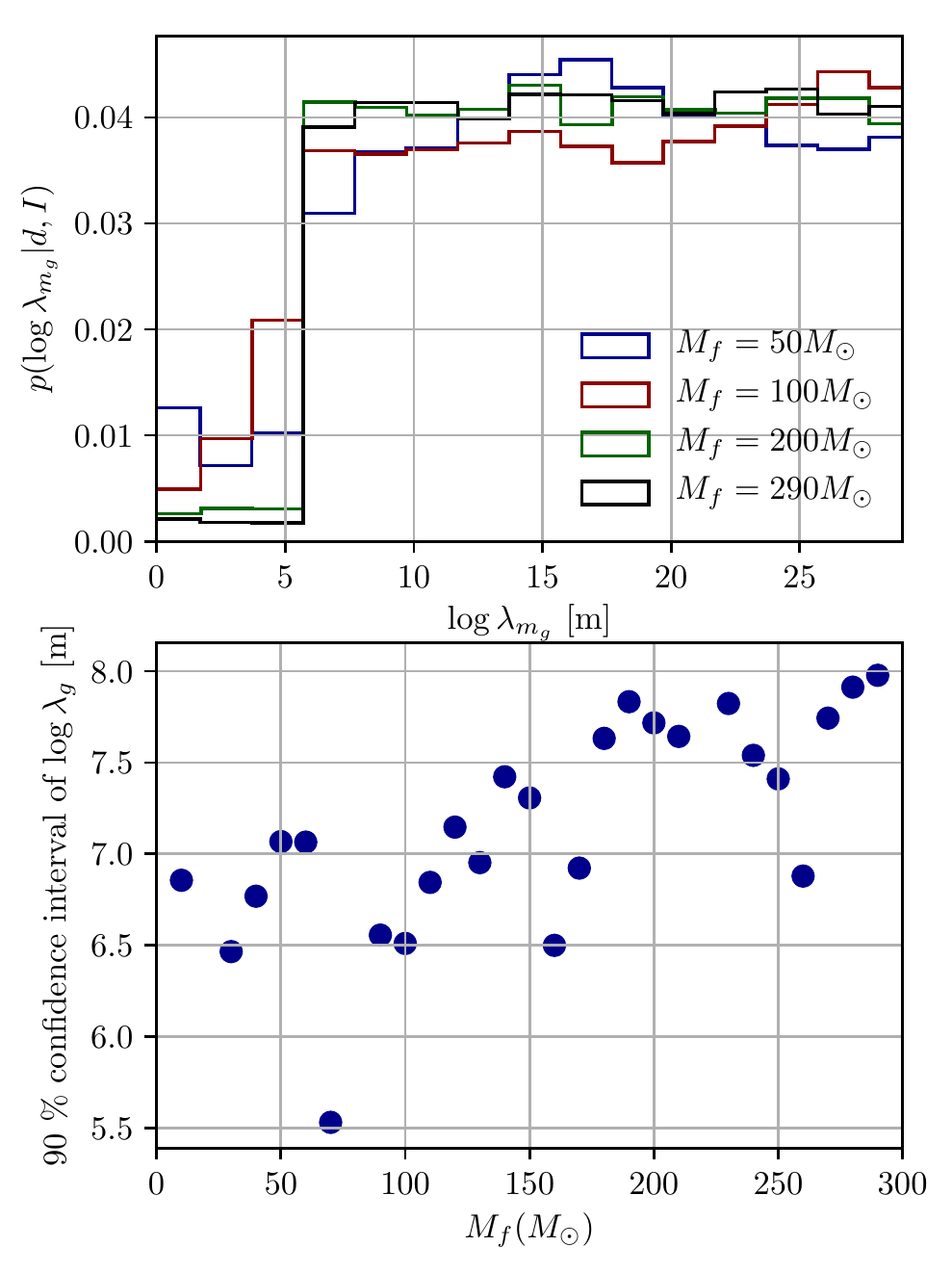}
  \caption{(Top panel) The posterior of $ \log \lambda_g $ obtained from the ringdown signal by a black hole of $ 50 M_{\odot}$ (blue), $ 100 M_{\odot}$ (red), $ 200 M_{\odot}$ (green) and $ 290 M_{\odot}$ (black) at 400\,Mpc. The posterior of different black hole masses are in step-function shape. Beyond $ \log \lambda_g \sim 6 - 8 $, the ringdown waveforms of $m_g > 0 $ are almost indistinguishable from $ m_g = 0 $ to the sampler. (Bottom panel) The 90\% confidence interval of $ m_g $ as a function of the final black hole mass $ M_f $. The credible interval increases with the final mass of black holes, despite some fluctuation due to noise. Increase of signal-to-noise ratio with mass leads to a better constraint by a black hole of higher mass.}
  \label{fig:pdf}
\end{figure}
The top panel of \cref{fig:pdf} plots the posteriors $ p(\log \lambda_g|d, I) $, corresponding to the ringdown signals from black holes of $ 50 M_{\odot}$ (blue), $ 100 M_{\odot}$ (red), $ 200 M_{\odot}$ (green) and $ 290 M_{\odot}$ (black) at 400\,Mpc.
The posteriors are step functions, because the measurement rules out low values for $\log \lambda_g$ (high values of $m_g$), which would produce discernible effects on the waveform.
%Thus, samples of large $ \log \lambda_g $ are likely to be sampled. 
The 90\% confidence interval of this posterior for $ M_f = 50 M_{\odot} $ is around $ \log \lambda_g \approx 6.7 $, corresponding to a constraint of $m_g < 10^{-13}\,\textrm{eV} $. 
Beyond $ \log \lambda_g \sim 6 - 8 $ the two classes of waveforms are indistinguishable (see \cref{fig:RDTDWF}). 
The bottom panel of \cref{fig:pdf} plots the 90\% confidence interval of $ p(\log \lambda_g|d, I) $ as a function of the final mass of black hole.
For black hole masses in the range of $ 10 - 290\,M_{\odot} $, the 90\% confidence interval of $ \log \lambda_g$ spans $\sim 5.0 - 8.0$, which corresponds to constraints on $ m_g $ in the range of $ 10^{-12} - 10^{-15}\,\textrm{eV} $. 
At a fixed distance, the signal-to-noise ratio of the ringdown signal increases with mass, which naturally leads to a better constraint.
Nevertheless, the increasing trend shows fluctuation due to noise.
The accuracy is consistent with \cref{fig:RDTDWF}, which shows that as $ m_g \sim 10^{-13}\,\textrm{eV}$, the two families of waveforms overlap almost completely.

Our constraints are not as tight as those put by the phase difference of the inspiral waveforms even with comparable signal-to-noise ratio.
This is due to the difference in physical scales between these two methods.
The phase difference compares the Compton's wavelength of the graviton to the propagation distance (see, for example, Eq. (28) of \cite{Wills_01}).
For a binary black hole systems at 400\,Mpc, the expected constraint is $ \lambda_g >> \sqrt{D_L / f} \sim 10^{16} \rm m$ .
On the other hand, detection of the dispersionless ringdown waveforms implies $ \lambda_g >> {\omega^{(0)}_{nlm}}^{-1} \sim M \sim 10^5 \rm m $ for a black hole of $ \mathcal{O}(300)\, M_{\odot} $.
The different physical scales of the two methods results in different constraints.

\section{CONCLUDING REMARKS}
We have proposed a convenient phenomenological modification to the Teukolsky equation that includes different types of alternative dispersion relations of gravitational waves.
In particular, we focus on the case of massive gravitational fields.
We find that both the frequencies and lifetimes are increased by the graviton mass, which is consistent with the previous numerical studies \cite{Massive_Gravity_01, Massive_Vector_Perturbation_01, Massive_Scalar_Perturbation_01, Massive_Scalar_Perturbation_02}.
These shifts leave signatures of the graviton mass in the ringdown waveform of black-hole merger system.
Compared to existing numerical studies, our work presents a simple analytical expression of quasinormal mode frequencies as a function of the graviton mass.
This makes inferring the graviton mass with black-hole ringdown signals possible in future studies.
Although this work concentrates solely about the graviton mass, it can be extended to more generic forms of Lorentz violation.

By including these shifts to an existing ringdown waveform model, we further demonstrated the ability to put constraints on the graviton mass solely by ringdown signals.
For black holes with masses in the range of $ 10 M_{\odot} $ to $ 290 M_{\odot} $ at $400\,\textrm{Mpc}$, observation of gravitational waves from the ringdown with an Advanced LIGO-Virgo network can constrain the graviton mass up to $ 10^{-12}\,\textrm{eV} $ to $ 10^{-15}\,\textrm{eV} $.
As it has been expected that the ringdowns of stellar mass black holes and their overtones can be detected \cite{Carullo_01, Brito_Buonanno_Raymond_01} by Advanced LIGO and Virgo at their design sensitivities \cite{LIGO_09}, our test of the graviton mass can be implemented for the future detection.
Constraints on the graviton mass can be improved by combining information from the inspiral and ringdown stages.

Although our constraint on the graviton mass are less stringent compared to those by inspiral waveforms, the latter concerns solely with the weak-field propagation of gravitational waves.
Instead, a test using the ringdown signal involves the strong-field regime.
Our studies shed light on the effects of alternative dispersion relations on the strong field generation of gravitational waves by relating the quasinormal mode frequencies of black holes to the graviton mass.

Lastly, our studies also provide additional insight into existing no-hair theorem tests.
The dependence of quasinormal mode frequencies on the dispersive properties of gravitational waves contradicts the no-hair theorem \cite{NHT_01, NHT_02, NHT_03, NHT_04}.
Existing tests of the no-hair theorem using gravitational-wave detections typically regard the fractional deviation of quasinormal-mode frequencies to be measurable free parameters \cite{NHT_TIGER_01, NHT_TIGER_02, NHT_TIGER_04, NHT_TIGER_03}.
These tests are model-independent but the physical meaning of these free parameters may not be immediately obvious.
Our works suggest that potential deviations of the quasinormal mode frequencies can be interpreted from the perspective of different types of dispersion relations.

\textit{Acknowledgement -- }
The authors would like to acknowledge Michalis Agathos, Nathan Johnson-McDaniel,  Miok Park, Maurice van Putten, B.S. Sathyaprakash, Kenneth Young and Nicolas Yunes for their stimulating discussion.
AKWC would like to thank Gregorio Carullo for his introduction to the ringdown waveform package \cite{Nikhef_01}, Adrian K.~H. Lai, Peter T.~H. Pang, Amitjit Singh, Alexander M. Tanaka and Jacky H.~T. Yip. for their comments on the manuscript and Robin S.H. Yuen for his advice about programming.
The work described in this paper was partially supported by grants from the Research Grants Council of the Hong Kong (Project No. CUHK 24304317), the Croucher Foundation of Hong Kong, and the Research Committee of the Chinese University of Hong Kong.

\textit{Appendix: Calculation in LPT -- }
According to logarithmic perturbation theory, the leading order shift of a quasinormal mode frequency due to a small perturbation potential $ V^{(1)} $ is given by \cite{LPT_01, LPT_02, LPT_03, LPT_04}
\begin{equation}\label{eq:LPT_01}
\tilde{\omega}^{(1)} = \frac{1}{2 \tilde{\omega}^{(0)}} \frac{\braket{u|V^{(1)}|u}}{\braket{u|u}}, 
\end{equation}
where $ u $ is the quasinormal mode solution for $ V^{(1)} = 0 $. 
$ \braket{u|V^{(1)}|u} $ and $ \braket{u|u} $ are formally defined as \cite{LPT_02, LPT_04} 
\begin{equation}
\begin{split} \label{eq:inner_product_01}
\Braket{u|u} &= \int_{- \infty}^{+\infty} dr_* u^{2} = \int_{r_+}^{+\infty} dr  \frac{r^2 + a^2}{\Delta} u^2 , \\
\Braket{u|V^{(1)}|u} &= \int_{-\infty}^{\infty} dr_* V^{(1)} u^{2} = \int_{r_+}^{+\infty} dr \frac{r^2 + a^2}{\Delta} V^{(1)} u^2, 
\end{split}
\end{equation}
where $ r_* $ is defined by $ \frac{d}{d r_*} = \frac{\Delta}{r^2+a^2} \frac{d}{dr}$ and $ r_+ $ is the outer event horizon of the Kerr black hole. 
For \cref{eq:SE}, $ u $ is the solution for $ m_g = 0$ corresponding to quasinormal mode frequency $ \tilde{\omega}^{(0)} $ of Kerr black holes in general relativity. 

For a general $ V^{(1)} $, both $ \braket{u|V^{(1)}|u} $ and $ \braket{u|u} $ diverge, leaving \cref{eq:LPT_01} indeterminate and regularization is needed. 
However, if $ V^{(1)} = m_g^2 \frac{r^2 \Delta}{(r^2 + a^2)^2}$, then $V^{(1)} \rightarrow m_g^2 $ as $ r \rightarrow + \infty $. 
Therefore, one can introduce a cutoff $ r = \Lambda $ so that one can regard $ V^{(1)} (r \geq \Lambda) = m_g^2 $. 
$ \braket{u|V^{(1)}|u} $ and $ \braket{u|u} $ then consist of two parts: one from $ r \in [r_+, \Lambda] $ and one from $ r \in [\Lambda, + \infty) $.
If we let  
\begin{equation}
\mathcal{I} = \int_{\Lambda}^{+\infty} dr  \frac{r^2 + a^2}{\Delta} u^2, 
\end{equation}
then, we can write 
\begin{equation}
\begin{split} \label{eq:inner_product_02}
\Braket{u|u} &= \int_{r_+}^{\Lambda} dr  \frac{r^2 + a^2}{\Delta} u^2 + \mathcal{I} , \\
\Braket{u|V^{(1)}|u} &= m_g^2 \Big(  \int_{r_+}^{\Lambda} dr  \frac{r^2}{r^2 + a^2} u^2 + \mathcal{I} \Big). \\ 
\end{split}
\end{equation}
Since $ u $ grows exponentially at spatial infinity for quasinormal mode solutions, $ \mathcal{I} \rightarrow + \infty $ as $ r \rightarrow + \infty$. 
Thus, we assume $ \tilde{\omega}^{(1)} \rightarrow m_g^2 / 2 \tilde{\omega}_{nlm}^{(0)} $ in the far-field limit where we observe gravitational waves. 

We also perform numerical integration to confirm the above calculations. 
The radial Teukolsky equation can be solved analytically \cite{exact_sol_TE}, and its solution is given by  
\begin{widetext}
\begin{align}
\label{eq:exact_sol}
R(x) = e^{i \epsilon \kappa x} (-x)^{-s-i\frac{\epsilon+\tau}{2}} (1-x)^{i\frac{\epsilon-\tau}{2}} \times \sum_{n=-\infty}^{n=\infty} \alpha_{n}^{\nu} F(& n+\nu+1-i\tau,-n-\nu-i\tau; 1-s-i\epsilon-i\tau;x)
\end{align}
\end{widetext}
where $\epsilon = 2 M \omega $, $\kappa = \sqrt{1 - (a/M)^2} $, $ \tau = (\epsilon - m a) / \kappa $, $ x = \omega(r_+ - r) / \epsilon \kappa$, $ \nu = l + \mathcal{O}(\epsilon^2) $ is the renormalized angular momentum, 
\begin{equation}
\begin{split}
a_{n}^{\nu} =& \frac{i \epsilon \kappa(n+\nu+1+s+i\epsilon)(n+\nu+1+s-i\epsilon)}{(n+\nu+1)(2n+2\nu+3)} \\
&\times (n+\nu+1+i \tau) \\
\end{split}
\end{equation}
and $ F $ is the hypergeometric function. 
The analytic solution of $ u(x) $ can then be determined by the transformation of $ u = \Delta^{s/2} (r^2 + a^2)^{1/2} R $.
 
\begin{figure}[htp!]
  \centering
  \includegraphics[width=\columnwidth]{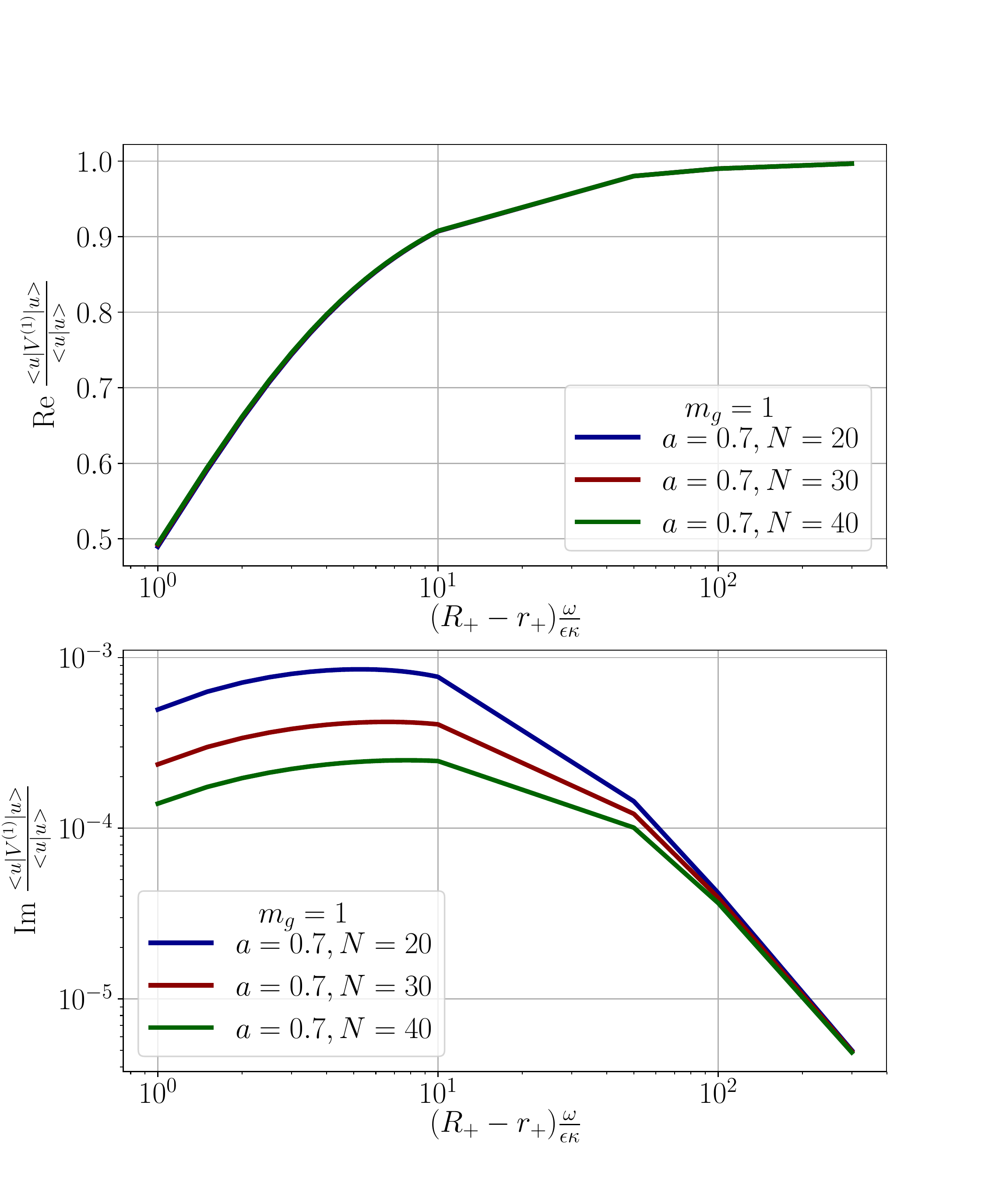}
  \caption{(color lines) The real (top panel) and imaginary part (bottom panel) of $ \Braket{u|V^{(1)}|u} / \Braket{u|u} $ as functions of the upper limit ($R_+$) of the inner products for $ m_g = 1 $ and $ a = 0.7 $. $ u $ is the analytic solution for the 022-quasinormal mode for Kerr black holes. $ u $ involves summation of hypergeometric function from $ n = - N $ to $ n = + N $. For $ N = 20, 30 $ and 40, the real part of $ \Braket{u|V^{(1)}|u} / \Braket{u|u} $ trends to 1 as $ R_+$ is extended to $ R_+ \sim r_+ + 50 \frac{\epsilon \kappa}{\omega} $. This serves a numerical proof of \cref{eq:freq_shift}}
  \label{fig:LPT}
\end{figure}

To numerically integrate \cref{eq:inner_product_01} using \cref{eq:exact_sol}, we pick $ \omega = \tilde{\omega}^{(0)} $, the 022-quasinormal mode frequency for Kerr black holes for $ a = 0.7 $. 
The values of $ M \tilde{\omega}^{(0)} $ for Kerr black holes of different $ a $ have been computed \cite{QNM_Kerr_01}.  
Due to computational limitation, it is impossible to include terms up to $ n = \pm \infty $. 
In practice, we sum the hypergeometric function in \cref{eq:exact_sol} from $ n = - N $ to $ n = + N $ for $ N =$ 20, 30 and 40.
Also, it is impossible to numerically evaluate the imporper integrals \cref{eq:inner_product_01}.
Instead, we evaluate the inner product on $ r \in [r_+, R_+] $ for some finite upper limit $ R_+ $. 
\cref{fig:LPT} plots the real (top panel) and imaginary part (bottom panel) of $ \Braket{u|V^{(1)}|u} / \Braket{u|u} $ for $ m_g = 1,  a = 0.7 $ and $ N = 20, 30 $ and 40 as functions of $ R_+ $. 
The horizontal axes denote $ R_+ - r_+ $, which, in principle, should be extended up to $ +\infty $. 
As \cref{fig:LPT} shows, for different values of $ N $, the real part of $ \Braket{u|V^{(1)}|u} / \Braket{u|u} $ trends to 1 as $ R_+$ is extended to $ \sim r_+ + 50 \frac{\epsilon \kappa}{\omega} $. 
Meanwhile, we also observe that the imaginary part of $  \Braket{u|V^{(1)}|u} / \Braket{u|u} $ trends to 0 as $R_+$ is extended to $ \sim r_+ + 50 \frac{\epsilon \kappa}{\omega} $. 
The numerical integration independently confirms that $ \tilde{\omega^{(1)}} \rightarrow m_g^2 / 2 \tilde{\omega}_{nlm}^{(0)} $ for the far-field limit. 

\bibliographystyle{apsrev4-1}
\bibliography{bibtex}

\end{document}